\documentstyle[twocolumn,amstex,aps]{revtex}
\begin{document}

\title{Spin Effects in the Local Density of States of GaAs}

\author{P. K\"onig, T. Schmidt, and R. J. Haug}

\address{Institut f\"ur Festk\"orperphysik, Universit\"at
Hannover, Appelstr. 2, 30167 Hannover, Germany}

\date{\today}
\maketitle

\begin{abstract}
We present spin-resolved measurements of the local density of states 
in Si doped GaAs. Both spin components exhibit strong mesoscopic fluctuations.
In the magnetic quantum limit, the main features 
of the spin-up and spin-down 
components of the local density of states 
are found to be identical apart from Zeeman splitting. 
Based on this observation, we introduce a mesoscopic method to measure 
the $g$-factor in a material where macroscopic methods are severely 
restricted by disorder. 
Differences between the spin-up and spin-down 
components
are discussed in 
terms of spin relaxation due to spin-orbit coupling.
\end{abstract}

Spin-polarized electronic transport is currently 
attracting a lot of interest
from both a fundamental and an applied point of view 
\cite{pri98}. 
Spin-polarized transport is essential
for the operation of spin transistors or spin valves, and recently
the spin injection from a ferromagnet into a semiconductor could be 
observed \cite{ham99}. 
Spin effects have also been proposed to be utilized for 
quantum computation. In this respect it is important to study
spin relaxation and coherence in semiconductors\cite{aws99}. 

Particularly interesting spin phenomena were observed in 
mesoscopic semiconductor 
structures. Prominent examples are the Kondo effect in 
quantum dots\cite{gol98,cro98} and spin-polarized tunneling 
through impurity levels.
Resonant tunneling through impurities
is also an established technique to image the local density of states (LDOS) 
of doped GaAs~\cite{sch96,sch97,des94,mcd95}. 
The LDOS exhibits mesoscopic fluctuations 
which can be understood in terms of interference of elastically 
scattered electron waves\cite{sch96,sch97}. 
However, a spin-resolved measurement 
was impossible to date.

In this paper, we investigate the spin dependence of the 
LDOS via resonant tunneling through a spin-split impurity level.
The strongest fluctuations of the spin-up and spin-down 
components of the LDOS are found to be identical apart from Zeeman splitting. 
This observation allows us to mesoscopically determine the $g$-factor 
in a material where macroscopic methods -- such as photoluminescence 
or magnetotransport -- are restricted by disorder. 
Differences between the spin-up and spin-down 
components are discussed in 
terms of spin relaxation due to spin-orbit coupling.

Our experiment is based on a strongly asymmetric double-barrier 
heterostructure, which was grown on $n^+$-type GaAs substrate.
It consists of a 10 nm wide GaAs quantum well and two Al$_{0.3}$Ga$_{0.7}$As 
barriers of 5 and 8 nm width. The nominally undoped active region is 
sandwiched between 300 nm thick GaAs contact layers doped with Si. The donor 
concentration has been experimentally determined to 
$3.3\times 10^{17}$ cm$^{-3}$ \cite{sch97}. From this material we fabricated 
a 2 $\mu$m diameter mesa with Ohmic contacts, which contains a small number 
of impurities in the quantum well.

We use the energetically-lowest impurity state
$S$ as spectrometer for the LDOS $\nu$ in the 
emitter contact adjacent to the thick barrier, see Fig.~\ref{sketch}. 
In a magnetic field $B$, the spectrometer exhibits a spin splitting 
$\Delta E_S=g_S\mu_BB$, with $g_S$ the $g$-factor of the impurity and 
$\mu_B$ the Bohr magneton. The spin splitting in the emitter is
$\Delta E=g\mu_BB$. By applying a bias voltage $V$, the spin-split levels 
$S_\uparrow$ and $S_\downarrow$ are shifted in energy with respect to the emitter. They cross the Fermi level at 
$V_{S_{\uparrow,\downarrow}}=(E_{S_{\uparrow,\downarrow}}-\mu_E^{ch})/e\alpha$
and two current steps result from resonant tunneling. 
Here, $E_{S_{\uparrow}}$ and $E_{S_{\downarrow}}$ are the energies of 
$S_\uparrow$ and $S_\downarrow$, $\mu^{ch}_E$ is the chemical potential in the 
emitter (measured from the band edge in the well and emitter, respectively), 
and $\alpha$ is the voltage-to-energy conversion coefficient. 
These two current steps can be clearly observed in Fig.~\ref{sketch}(a), 
which shows the current-voltage characteristic $I(V)$ of our device 
at the base temperature of $T=20$ mK of a dilution refrigerator. 
The spin splitting between the
steps increases as $B\parallel I$ is increased from 8.0 to 14.2 T. The overall 
shift of both steps to higher bias voltage is a consequence
of the diamagnetic shift of $E_S$ to higher energy. 

As the tunneling rate of the thick emitter barrier is much lower 
than that of the collector barrier, 
the tunneling current images the LDOS in the emitter below the Fermi level 
$\mu_E$ at the position of $S$\cite{sch96}. 
If spin is conserved during tunneling, the current on the 
short plateau of the first step is proportional to the spin-up 
component of the LDOS, $I_1\propto \nu_\uparrow$, while the current on 
the plateau of the second step reflects a superposition of both 
spin components, $I_2\propto \nu_\uparrow+\nu_\downarrow$. 
At a bias voltage of
$V_{\uparrow,\downarrow}=(E_{S_{\uparrow,\downarrow}}-E)/e\alpha$,
the spin-up and spin-down components of the LDOS at energy $E$ are 
probed by our experiment (measured from the band edge in the emitter).
The oscillatory fine structure on the current plateaus in 
Fig.~\ref{sketch}(a) reflects LDOS fluctuations in the heavily doped 
emitter \cite{sch96}.
These fluctuations are more pronounced in the numerically obtained 
differential conductance $G=dI/dV$ shown in Fig.~\ref{sketch}(b), where 
sharp peaks result from the current steps in the $I(V)$ data. 

To determine the $g$-factor of the spectrometer, we plot 
$\Delta V_S=V_{S_\downarrow}-V_{S_\uparrow}$ as a function of magnetic 
field. Figure~\ref{spins} 
shows that the data follows a straight line, which confirms that the two 
steps in the $I(V)$ curves reflect one spin-split impurity level. 
From the temperature dependence of the current step at 
$B=0$ T (not shown), we deduced a voltage-to-energy 
conversion coefficient of $\alpha=0.50$ for our device\cite{des96,su92,gei94}. 
Using this value, we obtain $|g_S|=e\alpha(d\Delta V_S/dB)/\mu_B=0.14$. 
The $g$-factor of the impurity is different from the bulk value
due to a change of the electron energy as a result of the 
double-barrier confinement and the penetration of the electron wavefunction 
into the barrier material\cite{ivc92}.  
Note that the spin splitting of the spectrometer does not tell whether 
$g_S$ is positive or negative. 

Now we turn our attention to the LDOS in the magnetic quantum limit,
in which only one spin-split Landau band remains populated in the emitter.
The transition to this limit occurs at $B\sim 11.4$ T. 
Figure \ref{cult}(a) shows a color-scale image of the current
as a function of bias voltage and magnetic field. It is dominated by 
two sharp current changes denoted by $V_{S_\uparrow}$ and 
$V_{S_\downarrow}$, which reflect the spin-split steps in the 
$I(V)$ curves in Fig. \ref{sketch}(a). 
The magnetic-field dependence of $V_{S_\uparrow}$
has been eliminated by shifting the $I(V)$ curves in voltage. 
Between $V_{S_\uparrow}$ and $V_{S_\downarrow}$ as well as above 
$V_{S_\downarrow}$, we observe pronounced fan-type resonances which 
reflect the local structure of Landau bands in the heavily doped
emitter \cite{sch97}. These resonances in the LDOS are formed by 
quasi-one-dimensional interference of multiply-scattered electron waves 
subject to Landau quantization. The slope of the LDOS 
resonances is $dV_E/dB=3.0$ mV/T\cite{foot2}.

To study the spin dependence of the LDOS, we define two 
cuts at $V_1$ and $V_2$ [dashed white lines in Fig.~\ref{cult}(a)]. 
Figure~\ref{cult}(b) shows the corresponding
currents $I_1$ and $I_2$ as a function of magnetic field. 
Assuming in a first attempt that the spin-up and 
spin-down components of the LDOS in the emitter 
are identical apart from a Zeeman shift
$\Delta E$, we can recalculate $I_2(B)$ from $I_1(B)$ as follows. 
From the sketch in Fig.~\ref{recon}, we deduce
\begin{equation}
I_1(B) \propto \nu_\uparrow(E_1,B) 
\label{i1}
\end{equation}
and 
\begin{equation}
I_2(B) \propto \nu_\uparrow(E_2,B) + 
\nu_\downarrow(E_2+\Delta E_S,B)
\label{i2}
\end{equation}
where $E_1$ and $E_2$ correspond to $V_1$ and $V_2$ and $\Delta E_S$ 
denotes the spin splitting of the spectrometer. 
Using $\nu_\downarrow(E,B) = \nu_\uparrow(E-\Delta E,B)$, we replace the
spin-down component of the LDOS in Eq.~(\ref{i2}) by its spin-up 
counterpart. In addition, we exploit the linear, fan-type character
of the LDOS resonances and use
$\nu_\uparrow(E_1+E,B) = \nu_\uparrow(E_1,B+[dB/dE_E]E)$
to rescale all energy values to $E_1$. 
Defining $E_{21}=E_2-E_1$, we obtain
\begin{equation}
\begin{split}
I_2(B) &\propto \nu_\uparrow\left(E_1,B+\frac{dB}{dE_E}E_{21}\right) \\
&+ \nu_\uparrow\left(E_1,B+\frac{dB}{dE_E}[E_{21}+\Delta E_S-\Delta E]\right) 
\\
\end{split}
\label{eb}
\end{equation}
from Eq.~(\ref{i2}). 
Using  Eq.~(\ref{i1}), we now rewrite Eq.~(\ref{eb}) into
\begin{equation}
\begin{split}
I_2(B) &= \gamma I_1\left(B+\frac{dB}{dE_E}E_{21}\right) \\
&+ \gamma I_1\left(B+\frac{dB}{dE_E}[E_{21}+\Delta E_S-\Delta E]\right)\;\;.\\
\end{split}
\label{fin}
\end{equation}
Here, $\gamma$ was introduced to compensate a difference in 
the coefficients of proportionality in Eqs.(\ref{i1}) and (\ref{i2}). 
It takes into account that the impurity can only be occupied by one 
electron at a time due to Coulomb interaction\cite{des96}.
For our analysis, 
we estimate $\gamma$ as $\langle I_2\rangle_B/2\langle I_1\rangle_B=0.98$ 
where $\langle ...\rangle_B$ symbolizes averaging over magnetic field.
The slope $dB/dE_E=1/[e\alpha (dV_E/dB)]=0.67$~T/meV 
and $|g_S|=0.14$ have been measured and, in a first step, 
we assume $g=-0.44$~\cite{wei77}.

Figure~\ref{recon} compares the experimental data for $I_2$ with 
$I_2^R$ recalculated according to Eq.~(\ref{fin}). 
Assuming in addition $g_S=-0.14$, the 
strongest features in the LDOS are nicely 
reproduced by our reconstruction. 
Thus we conclude that the spin dependence of large fluctuations in the 
LDOS can be understood in terms of Zeeman splitting. 
These features vary slowly in magnetic field (on a scale of about 
0.5 T or larger). The agreement also corroborates that both the spin-up 
and spin-down components of the LDOS are measured at the 
same location. The reconstruction does not work for $g_S=+0.14$, 
which strongly suggests that $g_S<0$ as predicted by calculations of 
the $g$-factor in a 10 nm wide quantum well~\cite{ivc92,tho98}. 
Strong deviations between $I_2$ and $I_2^R$ are observed for features that 
vary rapidly in magnetic field (on a scale shorter than 0.1 T). 
This is more evident at the bottom of Fig.~\ref{recon}, 
where we plot $\Delta I_2=I_2-I_2^R$ vs magnetic field. 

In order to determine the actual $g$-factor in the heavily doped 
GaAs emitter, we minimize the root-mean-square deviation 
$\sigma_{\Delta I_2}=\langle(\Delta I_2(B)-\langle\Delta I_2\rangle_B)^2 \rangle_B^{1/2}$ 
with respect to $g$. 
Figure~\ref{spinldos} shows a clear minimum at 
$g=-0.35$.
The accuracy of $g$ is limited by the uncertainty of 
$g_S$. Assuming an error of $\pm 0.01$ in $g_S$, we calculate the 
error of $g$ to be smaller than $\pm 0.02$. 
In our experiment, the $g$-factor is measured at the Fermi energy of 
about $25$ meV above the band edge and at magnetic fields 
of $10.5-15$ T, while the standard value 
$g=-0.44$ was obtained at the band edge and at 
zero field \cite{wei77}. The observed reduction of $|g|$ is in 
accordance with theory predicting a reduction 
of the $g$-factor in high magnetic fields and at finite 
electron densities \cite{bra85}.

Now we address the origin of the difference between 
$I_2$ and $I_2^R$ in Fig.~\ref{recon}. 
One issue to be considered is energy relaxation. 
The ``holes'' generated by tunneling in the emitter Fermi 
sea have an energy deficit $e\alpha (V-V_{S_{\uparrow,\downarrow}})$
with respect to $\mu_E$ (see sketch in 
Fig.~\ref{sketch}). The higher the bias voltage the larger is 
this energy deficit, resulting in a faster relaxation via 
electron-electron scattering. Inelastic scattering reduces the 
phase-coherence length in the emitter and thus suppresses the LDOS 
fluctuations. This effect is clearly visible in Fig.~\ref{sketch}, where 
the fluctuations become less pronounced with increasing bias voltage.
Similarly one might expect that $I_2$ is different from $I_2^R$, 
since $I_2$ is measured deeper below the Fermi level than $I_1$ which is 
used to calculate $I_2^R$. To check this idea, we calculated 
$\Delta I_2=I_2-I_2^R$ for 0.5~mV $\leq V_2 \leq 1$~mV corresponding
to an energy range of 0.25 meV.
Interestingly, $\Delta I_2$ is fairly independent 
from $V_2$, which demonstrates
that energy relaxation is irrelevant in our experiment.

Thus the difference between $I_2$ and $I_2^R$ either indicates
an inaccuracy in our reconstruction -- possibly resulting from 
a magnetic-field dependence of the parameters $\gamma$, $dB/dE_E$, 
$g_S$, $g$ -- or it gives evidence of additional spin effects 
beyond Zeeman splitting.
While we will not be able to give a definite answer, it is worth while 
to examine the influence of extra spin effects on the LDOS.

One possibility is spin relaxation due to spin-orbit coupling in the 
emitter\cite{aws99}.
In this case, we can estimate the spin-relaxation time  
as follows. Semiclassically speaking,  
the LDOS fluctuations are
a result 
of quantum interference of diffusive electron waves at length scales
between the mean free path 
and the phase-coherence length \cite{sch96}. 
Long interference paths correspond to weak features that vary rapidly in 
magnetic field. By calculating the differential conductance $G=dI/dV$, we 
favor small and rapid variations resulting from
long interference paths [see Fig.~\ref{sketch}(b)]. Due to the 
limited energy resolution $\hbar\Gamma$ of our spectrometer, the longest 
and dominant length scale in the conductance is 
$l_\Gamma=\sqrt{D/\Gamma}$, 
where $D$ is the diffusion constant 
in the emitter \cite{sch96,fal97}. 
The correlation magnetic field 
of the conductance fluctuations is
\begin{equation}
B_{c,G}\sim\phi_0/l_\Gamma^2=
\phi_0/(D/\Gamma) \;\;.
\label{bg}
\end{equation}
By calculating the difference $\Delta I_2=I_2-I_2^R$, we eliminate the
spin-independent features of the LDOS. Hence the correlation 
magnetic field of the fluctuations in $\Delta I_2$ contains 
information about spin relaxation,
\begin{equation}
B_{c,\Delta I_2}\sim\phi_0/l_{s}^2=\phi_0/D\tau_{s} \;\;,
\label{bdi2}
\end{equation}
with $l_{s}=\sqrt{D\tau_{s}}$ being 
the spin-relaxation length.
From Eqs.~(\ref{bg}) and (\ref{bdi2}), 
we now calculate the 
spin-relaxation time 
\begin{equation}
\tau_{s}\sim (B_{c,G}/B_{c,\Delta I_2})/\Gamma \;\;.
\label{tau} 
\end{equation}
All parameters in this expression have been measured in our experiment. 
In particular, it does not depend on the diffusion constant.
From the measured $G(B)$ and $\Delta I_2(B)$ data
we find $B_{c,G}\sim 0.01$~T and 
$B_{c,\Delta I_2}\sim 0.03$~T (in analogy to Ref. \onlinecite{sch96}), 
and from the width of the conductance peaks in Fig.~\ref{sketch}(b) 
we deduce an energy resolution of 
$\hbar\Gamma \approx e\alpha \times 80$~$\mu$V $\approx 40$~$\mu$eV.
Using these values, we obtain a spin-relaxation time of $\tau_{s}\sim 
5$~ps. 
It is much longer than the elastic scattering time $\tau=0.1$--0.2~ps, 
which we determined according to Ref. \onlinecite{fal97}.

Another interesting spin effect may result from 
the presence of magnetic impurities in the emitter. 
The exchange field of fully spin-polarized impurities 
leads to different disorder potentials 
for spin-up and spin-down electrons and thus influences 
the LDOS\cite{flip}. As there may only be residual magnetic impurities, 
their density will be by orders of magnitude lower than
the Si concentration. Therefore, the spin-dependent features of the 
disorder potential vary on longer length scales than the 
spin-independent features.
Slow variations in space correspond to small features on
short field scales in the LDOS.
Thus a small amount of magnetic impurities
could explain why small features in the LDOS exhibit an 
irregular spin dependence, while large features simply show a 
regular Zeeman splitting. To clarify the influence of
magnetic impurities on the LDOS,
we suggest to deliberately dope the GaAs
contacts of a double-barrier heterostructure with 
magnetic impurities such as Mn\cite{ohn98}. 

In summary, we used a spin-split impurity level as spectrometer to study 
the interplay of Zeeman splitting and disorder in Si doped 
GaAs via resonant tunneling. 
Both the spin-up and spin-down components of the LDOS 
show fan-type resonances  
as a consequence of Landau quantization in the bulk semiconductor. 
While the main resonances in the LDOS exhibit a Zeeman 
shift that can be best described using a $g$-factor of $g=-0.35$, 
the smaller resonances show a complicated irregular behavior.
Spin relaxation due to spin-orbit coupling is one possible 
origin of this deviation from Zeeman splitting. 

We gratefully acknowledge A. F\"orster and H. L\"uth for growing
the double-barrier heterostructure and V.I. Fal'ko for useful
discussions.

\begin{figure}
\caption{(a) Current-voltage curves $I(V)$ measured at 
$B=8.0$ T (solid) and 14.2 T (dashed) at $T=20$ mK. 
(b) Differential conductance $G=dI/dV$ numerically obtained from 
the $I(V)$ curves. 
The sketch shows the spin-resolved local density of states (left) 
in the emitter of a double-barrier heterostructure (shown to the right).
\label{sketch}}
\end{figure}
\begin{figure}
\caption{Spin splitting $\Delta V_S= V_{S_\downarrow}-V_{S_\uparrow}$ 
of the spectrometer vs magnetic field. 
\label{spins}}
\end{figure}
\begin{figure}
\caption{(Color). (a) Color map of the current
vs bias voltage (step 7~$\mu$V) and magnetic field 
$B\parallel I$ (step~10~mT) for $T=20$~mK 
(white,~$I\leq 0.0$~nA; black,~$I\geq 0.5$~nA).
The raw data has been shifted in voltage such that the
first current step defines origin of the voltage scale. 
The image is dominated by fan-type resonances $V_E$. 
(b) Current vs magnetic field at $V_1$ (dashed, $I_1$) and $V_2$ 
(solid, $I_2$).  
\label{cult}}
\end{figure}
\begin{figure}
\caption{
Reconstruction of $I_2$ from $I_1$: Comparison of the measured 
current $I_2$ (solid) with the reconstructed current $I_2^R$ (dashed). 
The solid line at the bottom shows the 
difference $\Delta I_2=I_2-I_2^R$ as a function of magnetic field. 
The sketch shows the local density of states (left) 
and the spectrometer position for $V_1$ and $V_2$ (right) to 
illustrate the reconstruction principle. Annotations are described 
in the text. 
\label{recon}}
\end{figure}
\begin{figure}
\caption{
Root-mean-square deviation $\sigma_{\Delta I_2}$ of $\Delta I_2=I_2-I_2^R$
as a function of the $g$-factor in the emitter.
\label{spinldos}}
\end{figure}


\begin{references}
\bibitem{foot2}In Fig.~\ref{cult}(a), the slopes of the spin-up and 
spin-down resonances are slightly different due 
to the spin splitting of the impurity level.
However, this effect is negligible
[$d\Delta V_S/dB=0.02$~mV/T $\ll dV_E/dB=3.1$~mV/T].
\bibitem{flip}At the low temperatures of our experiment, any magnetic 
impurities will be fully spin-polarized such that spin-flip scattering 
of electrons at these impurities is unlikely.
\end{references}
\end{document}